\documentclass[runningheads]{llncs}

\def\participants{154}
\def\messages{3,759}
\usepackage[T1]{fontenc}

\usepackage{subcaption}
\usepackage{graphicx}
\usepackage{multirow}
\usepackage{booktabs}
\usepackage{tikz}
\usepackage{hyperref}
\usetikzlibrary{arrows.meta, automata,
                calc,
                positioning,
                quotes}
\usepackage{csquotes}
\usepackage{enumitem}

\usepackage[english]{babel}
\addto\extrasenglish{%

}

\begin{document}

\title{A Human-Centric Evaluation of a Retrieval-Augmented Generation System for Explaining Quebec Insurance Contracts}
\titlerunning{A Human-Centric Evaluation of a RAG System for Explaining Insurance}

\author{David Beauchemin\inst{1}\orcidID{0000-0002-4084-8239}\thanks{Corresponding author.} \and
Richard Khoury\inst{1}\orcidID{0000-0001-8237-4548}}

\institute{GRAIL, Université Laval, Quebec, Canada\\
\email{\{david.beauchemin,richard.khoury\}@ift.ulaval.ca}
}

\maketitle

\begin{abstract}
With the rise of online insurance sales, consumers face a significant \enquote{advice gap}, requiring them to navigate complex legal contracts without expert guidance. 
This paper presents a human-centric, extrinsic evaluation of a state-of-the-art Retrieval-Augmented Generation system, designed to make Quebec automobile insurance contracts more understandable.
Through a user study with \participants{} participants from Laval University, we assess the agent's real-world utility by measuring system satisfaction, cognitive effort, perceived autonomy, and risk. 
Our results show the system is perceived as a \enquote{cognitive equalizer}, receiving high ratings for satisfaction, trust, and clarity. 
Crucially, users value the sense of autonomy the system provides even more than the knowledge itself, with this effect being most pronounced among participants with lower financial literacy, demonstrating how such an agent can directly empower individuals. 
However, our study also highlights some limitations: participants expressed a strong preference for human agents in high-stakes or emotionally charged scenarios.
This work highlights the importance of human-in-the-loop frameworks in ensuring the responsible implementation of AI in high-stakes consumer finance.
\end{abstract}

\section{Introduction}

Thanks to recent changes in the legal landscape, such as the 2019 implementation of Bill 141 in Quebec, insurers can now sell products entirely online without the intervention of a human insurance representative \cite{rlrq2004}. 
This legislative shift, while creating new commercial opportunities, introduces a significant \enquote{advice gap} for consumers \cite{protectionepargnants}. 
Customers are now solely responsible for navigating complex insurance contracts filled with legal terminology, a daunting task given that insurance contracts are long and complex to read \cite{beauchemin-etal-2020-generating,cox2025legal}. 
This situation is compounded by the consumer's inherent vulnerability, stemming from an asymmetry of information, and by the fact that a portion of the population has a limited understanding of insurance products \cite{protectionepargnants}, or worse, is functionally illiterate \cite{alphabete}. 
This high-stakes environment requires innovative solutions to ensure consumers' safety in the digital marketplace \cite{amf2018memoire}.
Current methods for explaining insurance products are split between traditional and digital approaches. 
The traditional method relies on insurance representatives who provide personalized advice and tailored explanations. 
While effective, this model is not well-suited to an increasingly digital market \cite{cefrio}. 
Digital solutions encompass passive internet resources from regulatory bodies, such as the \textit{Autorité des marchés financiers} (AMF), as well as insurers' websites and FAQs. 
However, these online resources require users to actively search for information, and the simplified explanations they provide are the result of a lengthy manual creation process. 
For instance, a project by the AMF to create a new simplified version of the standard car insurance policy took nearly four years to finalize \cite{gaa_fpq1_2010-2014}. 
Consequently, current methods force a choice between the scale of digital platforms and the personalization of an expert agent, with no existing solution effectively combining both.

To address this gap, this paper investigates how a Retrieval-Augmented Generation (RAG) system can automatically provide accurate and understandable answers to customer questions. 
A RAG system enhances a Large Language Model's (LLM) responses by first retrieving relevant text from a specialized knowledge base, enabling it to generate an answer that is more accurate and factually grounded \cite{gao2023}. 
This approach has shown significant promise in complex domains such as law and insurance \cite{louis2024,beauchemin2024quebec}. 
However, the deployment of such agents introduces a critical challenge: balancing their potential to empower users against the inherent risks of AI-generated hallucination \cite{weidinger2021ethical,bender2021dangers}.
The primary objective of this research is to evaluate the real-world utility of such a system designed for this purpose.
Specifically, we investigate how the dynamics of human-agent interaction can bridge the \enquote{advice gap} and empower users in a high-stakes financial domain.
The main contribution of this paper is a comprehensive human evaluation of a RAG system's effectiveness, based on a study with \participants{} users. 
While the study's conclusions are robust for a more educated, high-literacy user group, caution should be exercised when generalizing these findings to the broader population of insurance consumers.
This article is structured as follows: \autoref{sec:related} details the related work, while \autoref{sec:methodology} details our methodology. 
\autoref{sec:results} presents the quantitative and qualitative results from the user study, and \autoref{sec:generatlization} presents our generalization findings. Finally, we conclude the paper and present our future work.

\section{Related Work}
\label{sec:related}
The application of LLMs to specialized domains, such as law and finance, is a rapidly growing area of research \cite{siino2025exploring}. 
In particular, RAG has emerged as a key architecture for improving the factual accuracy and domain-specificity of LLM-generated responses \cite{gao2023}. 
Recent work has demonstrated the effectiveness of RAG in legal question-answering \cite{beauchemin2024quebec}. 
These systems are designed to interpret statutory law, and new benchmarks, such as LegalBench-RAG, have been developed to evaluate performance on complex legal reasoning tasks \cite{pipitone2024}. 
The insurance sector, a specific type of legal and financial domain, has also seen initial applications, with chatbots developed to handle customer queries by leveraging domain-specific datasets \cite{nuruzzaman2020,beauchemin-etal-2020-generating}. 
While these studies establish the technical viability of RAG for domain-specific QA, our work shifts the focus from purely technical performance to a deep, human-centric evaluation of such systems in a consumer-facing role, assessing their real-world impacts on user understanding and autonomy.

The evaluation of user-facing AI systems has increasingly shifted away from automated, intrinsic metrics toward extrinsic, human-centric methodologies that measure real-world utility \cite{jurafsky2025}. 
The usability of chatbots is a developing field, with research focusing on establishing robust evaluation methods and understanding the nuances of user interactions with conversational interfaces instead of traditional graphical ones \cite{ren2022,nguyen2022}.
A critical factor in the adoption of AI agents in high-stakes domains is user trust. 
Studies have shown that users are more likely to trust AI systems that provide clear justifications for their outputs \cite{shin2021}. 
Our research builds on this body of work by conducting a large-scale extrinsic study within the Quebec (francophone) insurance context, providing empirical data on how factors such as \enquote{Cognitive Effort} contribute to the overall user experience (UX) of a specialized RAG.

\section{Methodology}
\label{sec:methodology}

To evaluate the RAG system's real-world utility, we designed and conducted an extrinsic, human-centric user study. 
Our methodology is grounded in established frameworks for human-computer interaction and chatbot evaluation, combining quantitative and qualitative analysis to provide a holistic assessment of the system's performance \cite{ren2022,barbosa2022ux}. 
This section details the system, participant recruitment, the UI, our data collection and analysis procedures, study sample and ethical considerations.

\subsubsection{System Architecture}
The insurance RAG system under evaluation utilizes OpenAI \enquote{o3-2025-04-16} LLM, along with the same RAG as described in \cite{beauchemin2024quebec}.

\subsubsection{Participant Recruitment and Profile}
Participants were recruited from the student and employee populations at Laval University.
The recruitment process involved a qualification survey designed to assess their insurance literacy. 
This survey consists of 15 agree/disagree questions adapted from established financial literacy studies \cite{tennyson2011,barbosa2022ux,morin2015,ren2022}. 
To contextualize our sample, participants' scores were benchmarked against the \enquote{Indice AMF}, an official measure of financial literacy in Quebec \cite{amf2018indice,cachecho2022} (see our supplementary materials). 

\subsubsection{User Interface}
To facilitate a controlled and realistic interaction, we developed a dedicated UI for the study, inspired by established research on chatbot usability and UX \cite{nguyen2022,yin2024}. It is illustrated in our supplementary materials.

\subsubsection{Data Collection}
Our data collection strategy mostly relies on a structured survey. 
Upon completing their 30-minute interaction with the chatbot, participants are directed to a post-use 21-question evaluation survey. 
This instrument is designed to assess the user's experience across four key themes derived from \cite{mcauley1989,task2008,cocca2022,nguyen2022,yin2024}: \enquote{System Satisfaction} (SS), \enquote{Cognitive Effort} (CE), \enquote{Perceived Autonomy} (PA), and \enquote{Perceived Risks} (PR). 
Our questionnaire is included in our supplementary materials. 
Most questions utilized a 7-point Likert scale \cite{Likert1932ATF}. 

\subsubsection{Statistical Procedures}
We employ a mixed-methods approach for data analysis. 
First, quantitative data from Likert-scale responses are analyzed using descriptive and inferential statistics. 
To compare the experiences of low- and high-literacy users, using a median split of their literacy scores, we use the independent-samples \textit{t}-test and the paired-samples \textit{t}-test to explore relationships across different dimensions.
All statistical tests use a significance level of $p < 0.01$.
Second, we conduct a correlation analysis using Pearson's correlation coefficient to better understand the relationships among the different dimensions.
Third, to reduce the potential for noise or bias inherent in any single question, we leverage a composite score (CS). 
This approach involves summing responses from multiple conceptually-related Likert-scale items into a single value. 
Fourth, we use inferential statistics to examine the relationship between insurance literacy and the utility of the RAG system. 
We hypothesize that the system's value may be perceived differently by individuals with varying degrees of literacy. 
To assess the statistical significance of any observed differences, given the ordinal nature of the Likert scale data, the Kruskal-Wallis (KW) H-test \cite{macfarland2016kruskal} non-parametric statistical test is employed.
Finally, to identify recurring patterns and extract nuanced insights from the qualitative data and open-ended comments, we use a thematic analysis approach.

\subsubsection{Study Sample}
Our user study successfully concluded with the collection of complete responses from \participants{} participants, along with the \messages{} message-answer pairs between participants and the RAG system, over a period of 30 days.

\subsubsection{Ethical Considerations}
The evaluation protocol and data collection tools were approved by the Université Laval Research Ethics Committee (2024-431, January 27, 2025). 
Respondents who completed the surveys consented to participating and to their answers being used for this project.

\section{Results and Discussion}
\label{sec:results}

\subsection{Quantitative Results}

\subsubsection{User Ratings}
\begin{figure}
        \begin{subfigure}[b]{0.32\linewidth}
            \centering
            \caption{\enquote{System Satisfaction} (SS) (part 1)}
            \label{fig:q1q5}
            \includegraphics[width=1\linewidth]{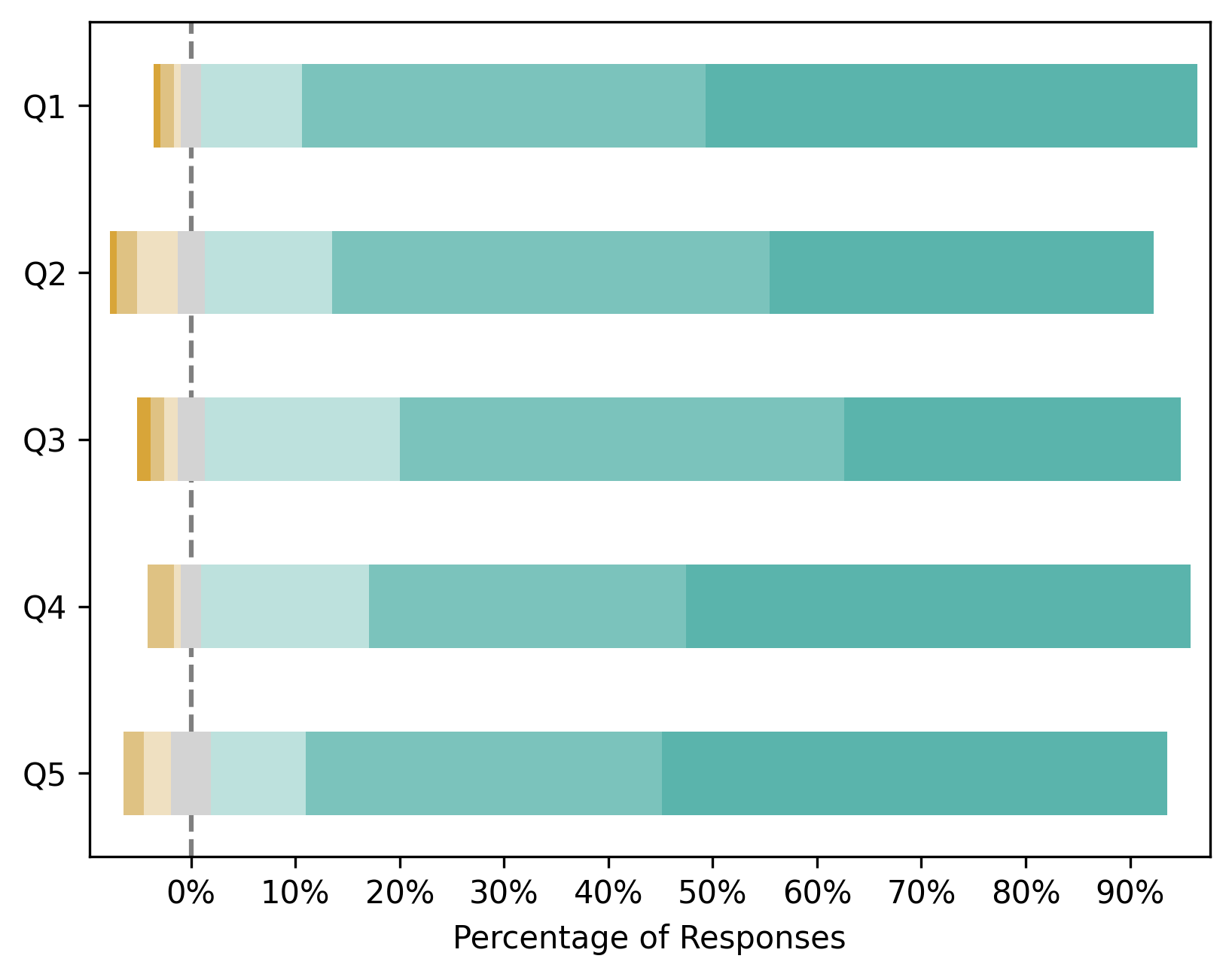}
        \end{subfigure}
        \begin{subfigure}[b]{0.32\linewidth}
            \centering
            \caption{\enquote{System Satisfaction} (SS) (part 2)}
            \label{fig:q1q10}
            \includegraphics[width=1\linewidth]{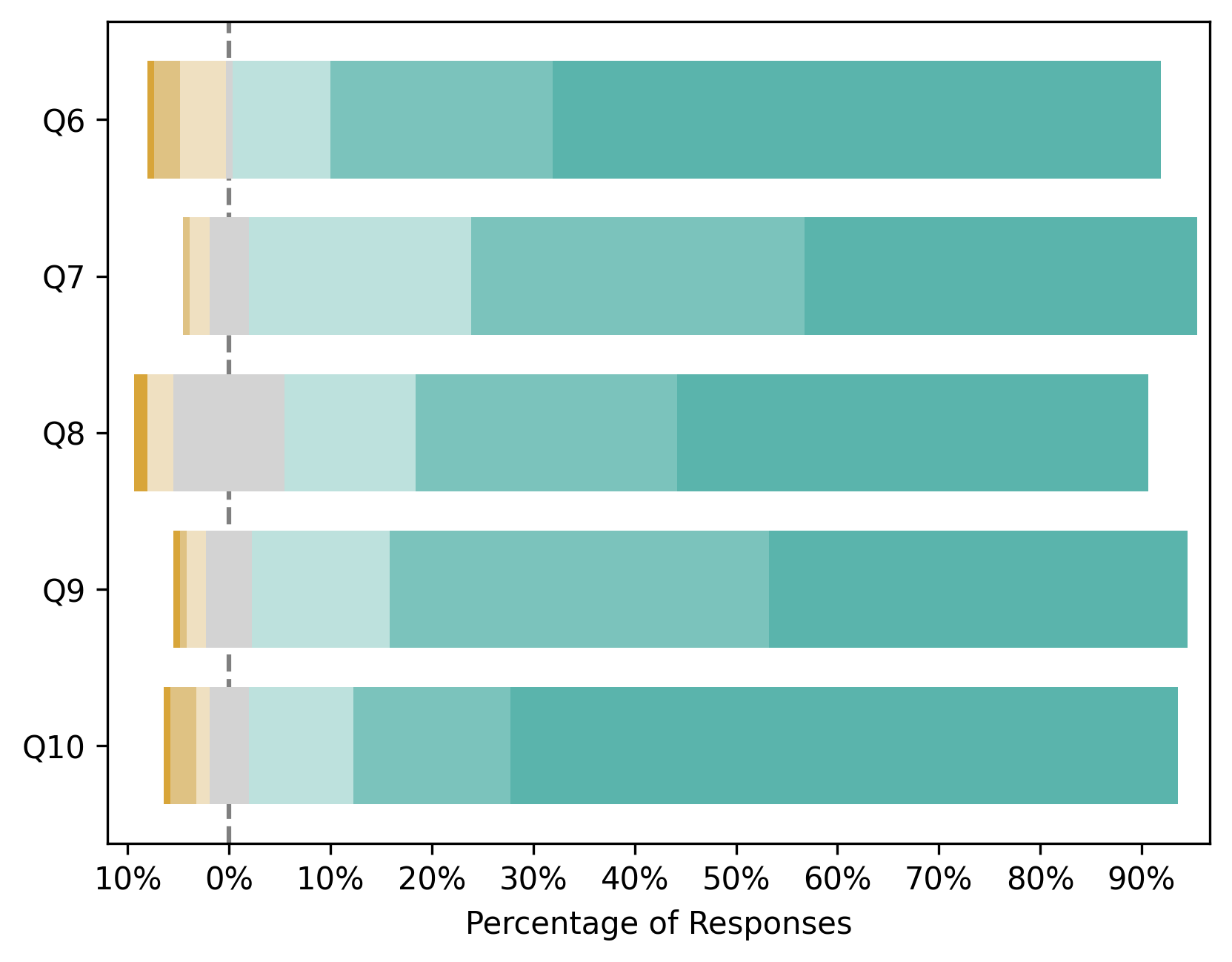}
        \end{subfigure}
        \begin{subfigure}[b]{0.32\linewidth}
            \centering
            \caption{\enquote{Cognitive Effort} (CE)}
            \label{fig:q12q15}
            \includegraphics[width=1\linewidth]{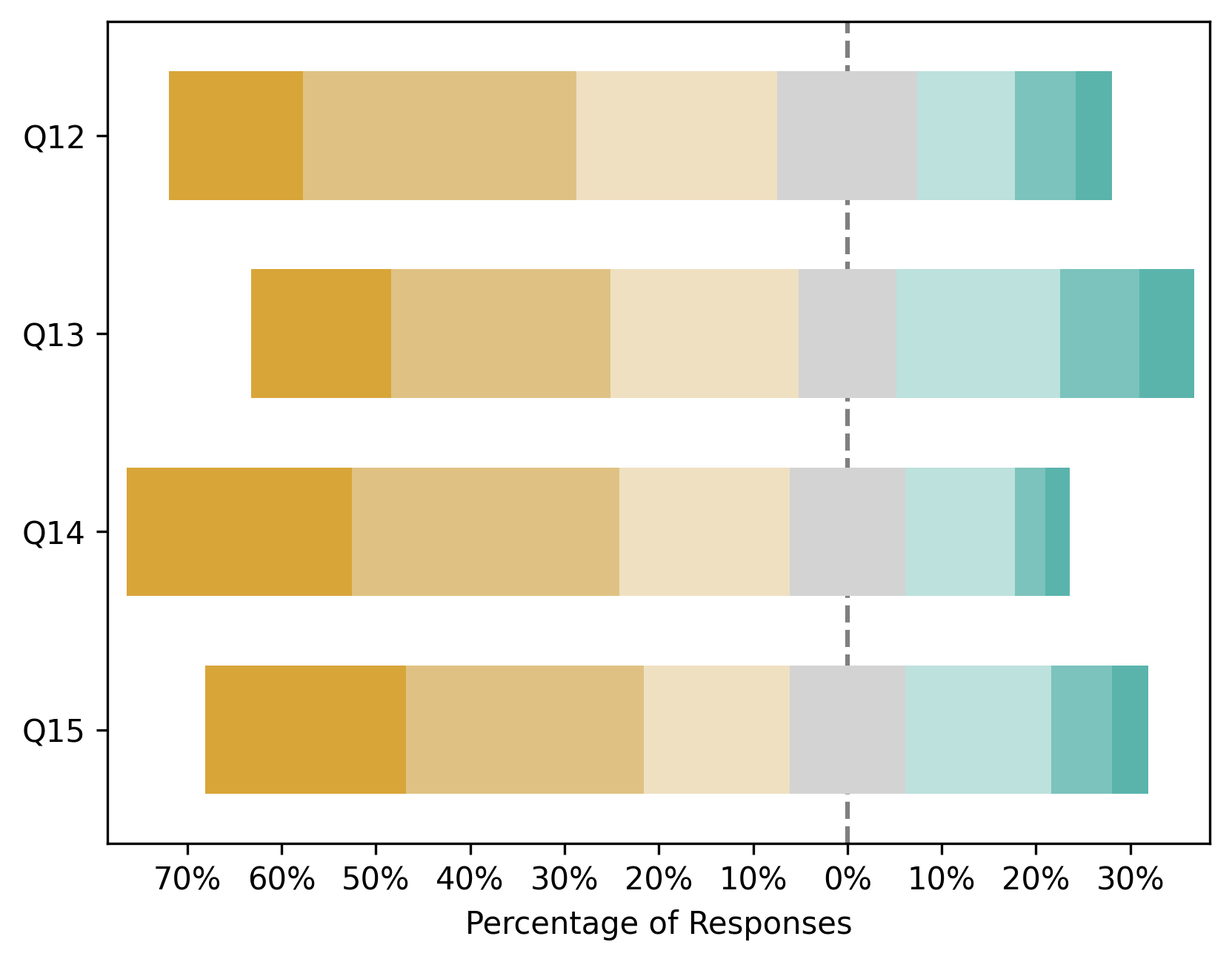}
        \end{subfigure}
        \begin{subfigure}[b]{0.32\linewidth}
            \centering
            \caption{\enquote{Perceived Autonomy} (PA)}
            \label{fig:q16q17}
            \includegraphics[width=1\linewidth]{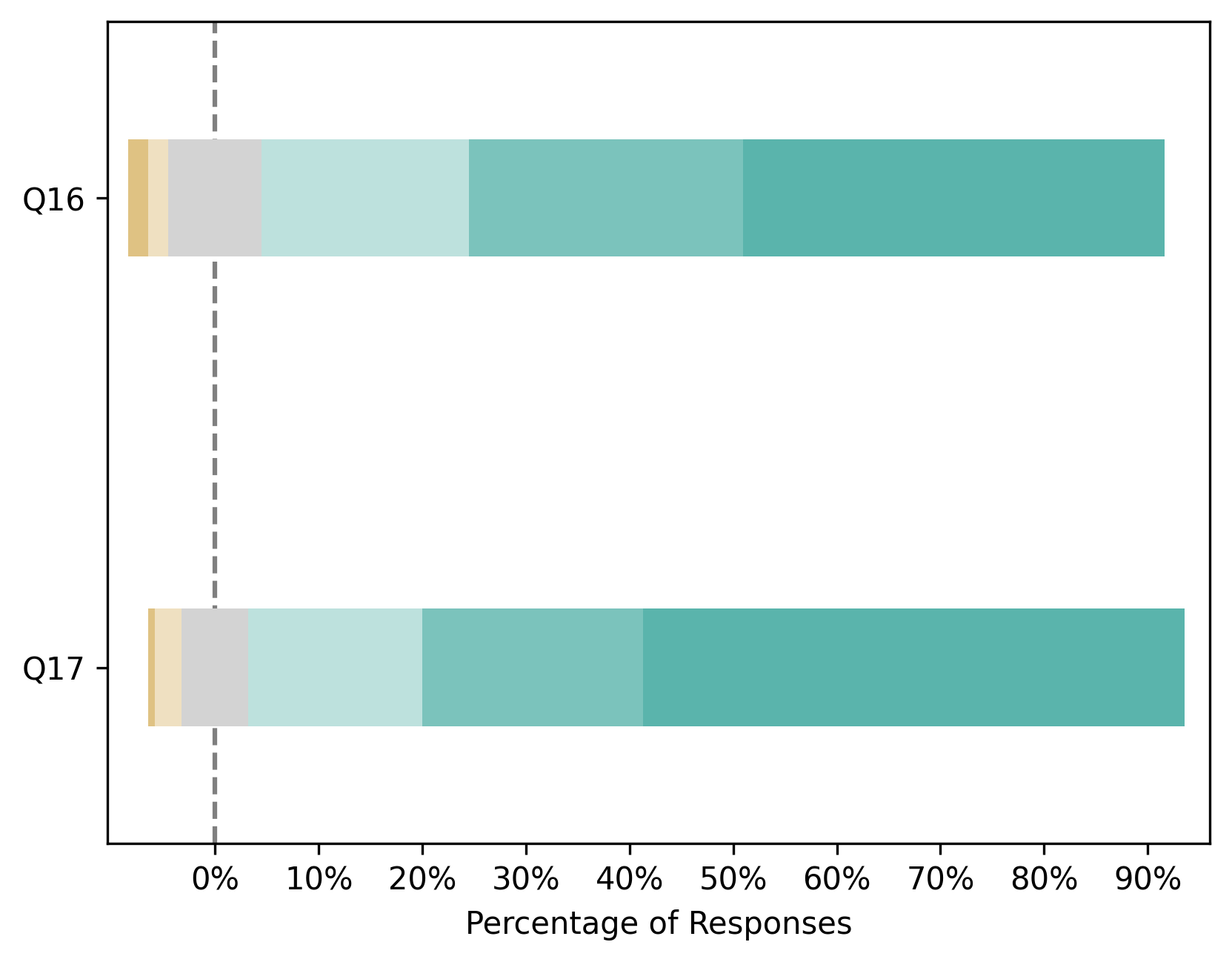}
        \end{subfigure}
        \begin{subfigure}[b]{0.32\linewidth}
            \centering
            \caption{\enquote{Perceived Risks} (PR)}
            \label{fig:q18q19}
            \includegraphics[width=1\linewidth]{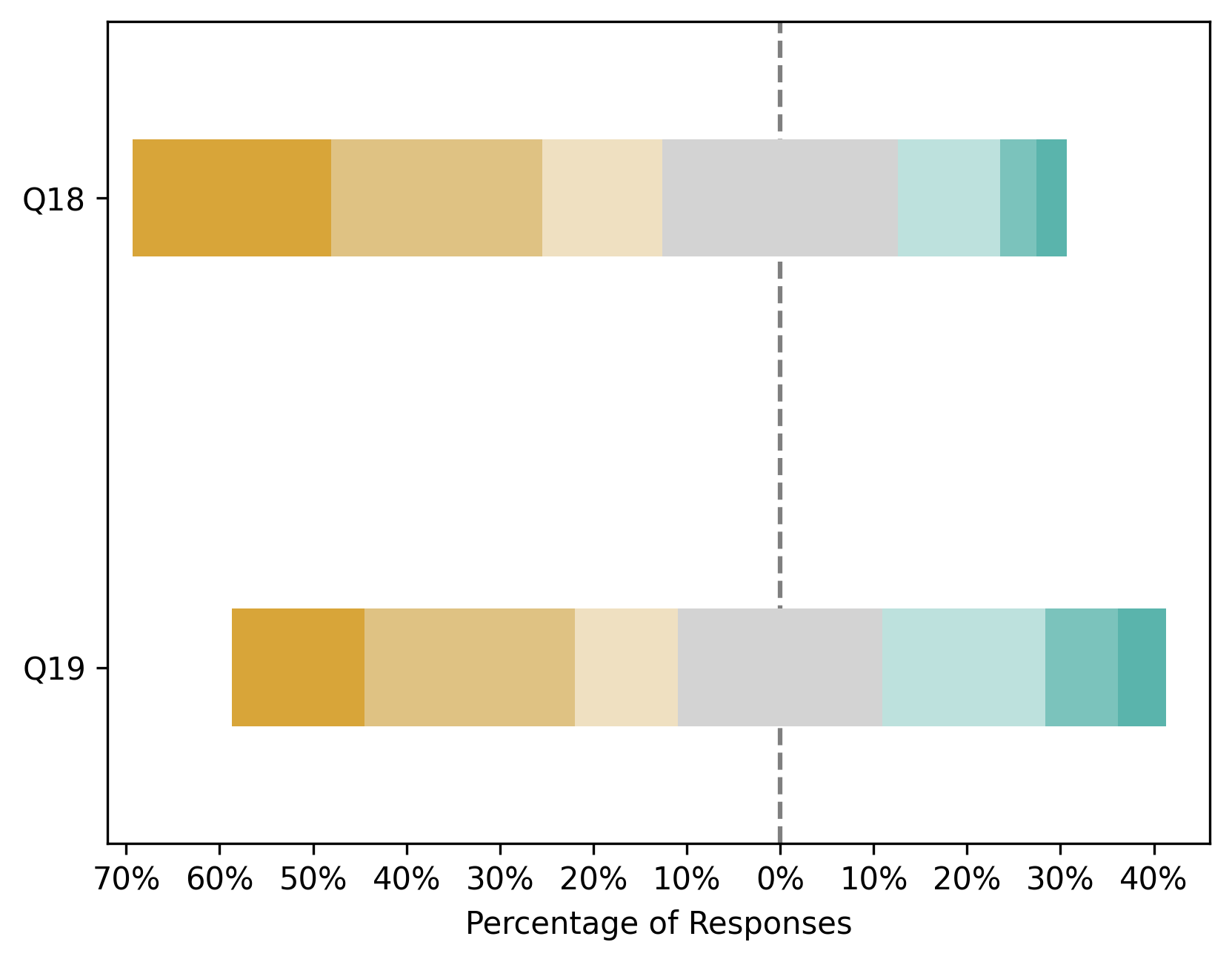}
        \end{subfigure}
        \begin{subfigure}[b]{0.32\linewidth}
            \centering
            \caption{Legend}
            \label{fig:legend}
            \includegraphics[width=\linewidth]{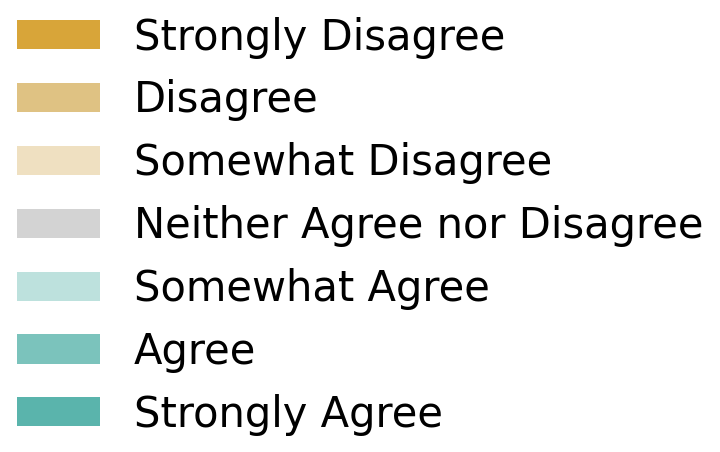}
        \end{subfigure}
    \caption{User feedback using a diverging stacked bar graph of our Likert scale. The dotted line refers to the neutral score (i.e. 4).}
    \label{fig:likertstacked}
\end{figure}

The quantitative data from the post-use evaluation survey reveal an overwhelmingly positive reception of the RAG system. 
As shown in the diverging stacked bar graphs of user responses to our Likert-scale questions (\autoref{fig:likertstacked}), sentiment was strongly skewed towards high agreement across all measures of system satisfaction (Q1-Q10). 
A visual inspection of the graphs for \enquote{SS} (\autoref{fig:q1q5} and \autoref{fig:q1q10}) highlights this trend, with the positive sentiment categories (\enquote{Agree} and \enquote{Strongly Agree}) dominating the response distribution for nearly every question, while negative responses are minimal.
Moreover, the graph for \enquote{CE} (Q12-15) (\autoref{fig:q12q15}) shows the opposite distribution; since these questions were framed negatively, the overwhelming disagreement visually confirms the system's low cognitive friction.
This visual skew across multiple themes underscores the robustness of the positive UX and shows that participants found the agent's responses clear and accessible, suggesting the system successfully functions as a \enquote{cognitive equalizer} that simplifies complex information.
This level of usability is further corroborated by the chatbot's responses `like/dislike' ratings. As shown in \autoref{tab:sentiment_analysis}, almost 82\% of the RAG's 3,759 responses were rated positively by users, while less than 3\% were rated negatively.

\begin{table}
    \centering
    \begin{tabular}{lcc}
    \toprule
             & Count & Frequency (\%) \\\midrule
    Positive & 3,080 & 81.93    \\
    Neutral  & 575   & 15.30     \\
    Negative & 104    & 2.77    \\\bottomrule
    \end{tabular}
    \caption{Results of the sentiment analysis on the message corpus.}
    \label{tab:sentiment_analysis}
    
\end{table}

\subsubsection{Correlations in User Experiences}
\begin{figure*}
    \centering
     \includegraphics[trim={2mm 2mm 2mm 2mm}, clip, scale=0.4, keepaspectratio]{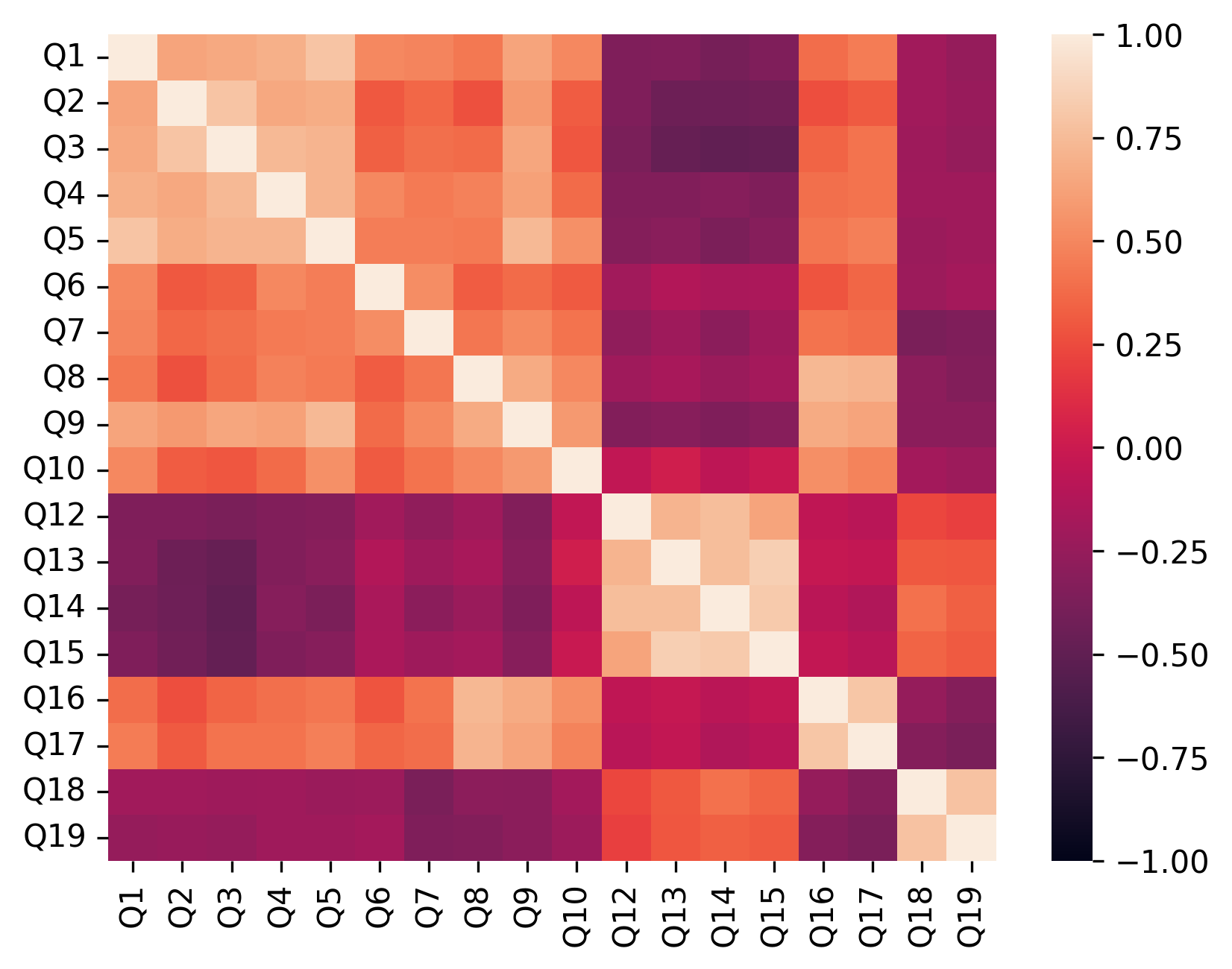}
    \caption{Pearson correlation heatmap matrix of user experience dimensions.}
    \label{fig:heatmap}
\end{figure*}

We present in \autoref{fig:heatmap} a Pearson correlation analysis of the questionnaire responses.
First, we can see that the intention to continue using the system (Q8) strongly correlates with personal autonomy (Q16) (0.73). 
It indicates that the feeling of empowerment is a primary motivator for long-term user engagement; users want to use the system precisely because it makes them feel more self-sufficient.
The set of questions related to satisfaction (Q1-Q5) is strongly negatively correlated with the block related to \enquote{CE} (Q12-Q15). It confirms that an effortless experience is a major contributor to user satisfaction. 
Furthermore, the analysis shows a moderate but significant negative correlation between trusting the agent's answers (Q7) and perceiving personal risk (Q18) (-0.38).
This relationship indicates that as users' trust in the system's reliability increases, their apprehension about potential negative consequences decreases, underscoring that building a trustworthy system is the most direct path to overcoming user hesitation.

\subsubsection{Inferential Statistics}
We present in \autoref{tab:inferal} the inference statistics for various questions and CS across the low- and high-literacy segments of our users.
First, we can see that the system's utility is most significant for participants with low literacy. 
Members of this group reported significantly higher overall satisfaction than their high-literacy counterparts, a finding confirmed by both parametric and non-parametric tests on individual satisfaction question Q1 and a CS(Q1-5, Q9). 
This result demonstrates the system's effectiveness as a \enquote{cognitive equalizer}, successfully bridging the knowledge gap for more vulnerable consumers. 
Interestingly, this difference was isolated to satisfaction; the levels of trust (Q7), perceived educational benefit (Q10), and empowerment (Q16) were statistically uniform across both groups. 
This nuance suggests that while novices derive greater satisfaction from the system, its core benefits are universally recognized and appreciated by all users, regardless of their literacy.

\begin{table}
    \centering
    \begin{tabular}{@{}lcc|cc@{}}
    \toprule
     & \textit{t}-Test & $p$-value    & \begin{tabular}[c]{@{}c@{}}Kruskal-Wallis H-test\end{tabular} & $p$-value\\ \midrule
     Q1 & -3.79  & $< .001$ & 8.28 & 0.004\\
     CS(Q1-5, Q9) & -4.72 & $< .001$ & 11.04 & $< .001$\\
     Q7 & 0.93  & 0.354 & 0.435 & 0.510\\
     Q10 & 1.22  & 0.225 & 6.54 & 0.011\\
     Q16 & 1.55  & 0.124 & 0.20 & 0.655\\
     \bottomrule
    \end{tabular}%
    \caption{Results of inferential tests between low/high-literacy groups.}
    \label{tab:inferal}

\end{table}

We present a set of paired-samples \textit{t}-tests in \autoref{tab:pairedttest}.
The first pair shows that, while participants agree that the agent improved their understanding (Q10), they rate its potential to increase their autonomy (Q16) significantly higher ($p < .001$). 
It indicates that users perceive the agent not merely as a passive source of information but as an active tool that enhances their capacity to manage their insurance needs. 
This feeling of empowerment, which is strongly correlated with the intention to use in the future (0.73 in \autoref{fig:heatmap}), appears to be the primary driver of the system's value proposition. It underlines a central finding of this study: users valued the system's ability to empower them more than its purely educational function. 
Second, this analysis reveals an interesting duality between the system's usefulness and its \enquote{PR} by users. 
Indeed, participants perceive this type of system as posing a significantly higher risk to society (Q19) than to themselves (Q18). 
However, their strong desire to continue using the system (Q8) robustly outweighed both their concerns about personal and societal risks, as confirmed by the highly significant \textit{t}-test results of both the Q8-Q18 and Q8-Q19 pairs. 
This duality between enthusiasm for the tool and risks of using it may explain the results of Q20, which asked users to choose which insurance tasks they would use the AI assistant for. As shown in \autoref{tab:q20}, a large majority of participants indicate a willingness to use the agent for informational and transactional tasks like purchasing a policy (76.81\%), while, on the other hand, they are more hesitant to perform more sensitive tasks like a claim (67.29\%). 

\begin{table}
    \centering
    \begin{tabular}{@{}lcc@{}}
    \toprule
      & Paired-Samples \textit{t}-Test & $p$-value    \\ \midrule
     CS(Q10-Q16) & -3.82  & $< .001$\\
     CS(Q18-Q19) & -3.93  & $< .001$\\
     CS(Q8-Q18) & -2.55  & 0.012\\
     CS(Q8-Q19) & -5.27 & $< .001$\\
     \bottomrule
    \end{tabular}
    \caption{Results of the paired-samples \textit{t}-test comparing the \textbf{same} participant response to different questions.}
    \label{tab:pairedttest}

\end{table}

\begin{table}
    \centering
    \begin{tabular}{lc}
        \toprule
         & \begin{tabular}[c]{@{}c@{}}\% of participants\end{tabular} \\
        \midrule
        Purchasing an auto and/or home insurance policy & 76.81 \\
        Making changes to my insurance file or policy & 75.36 \\
        Renewing insurance & 71.74 \\
        Making a claim following a loss & 67.29 \\
        Cancelling insurance & 56.52 \\
        None of the above & 10.87 \\
        \bottomrule
    \end{tabular}%
    \caption{Participants' answers to the multiple-choice question (Q20).}
    \label{tab:q20}

\end{table}

\subsection{Qualitative Results}
To complement our quantitative findings, we conducted a thematic analysis of open-ended feedback collected from the post-use evaluation survey, and four primary themes emerged.

\subsubsection{The System as an Empowering Ally}
The most prevalent theme is the system's ability to make complex insurance concepts understandable to users.
Participants commented that the chatbot \enquote{translated jargon into plain language} and \enquote{broke down complicated clauses into simple points}. 
It directly explains the low \enquote{CE} reported by users: the system performed the heavy lifting of simplifying complex legal terminology, allowing users to engage more confidently and thereby reducing the mental effort that can lead to poor decision-making.
For layperson users, this creates a powerful sense of psychological safety. 
One participant articulated this perfectly: \enquote{It's less intimidating than calling an agent. I can ask basic questions without feeling stupid.}
By removing this social friction, the chatbot fostered a judgment-free learning environment that lowered the barrier to seeking help, a critical factor in empowering users who feel vulnerable due to their lack of knowledge.

\subsubsection{From Passive Knowledge to Actionable Confidence}
User feedback indicates that the system's value lies in translating passive knowledge into a tangible sense of empowerment. 
Participants' comments consistently moved beyond \enquote{now I understand} to \enquote{now I know what to do}. 
As one participant commented, \enquote{It's one thing to read a definition of an endorsement. It's another thing entirely to feel confident enough to ask for a specific one when getting a quote. The agent gave me that confidence}.
This transition provides a compelling explanation for the quantitative finding that users valued the increase in their autonomy significantly more than the improvement in their understanding (Q10-16 in \autoref{tab:pairedttest}). 
This feeling of preparedness is a stark contrast to the \enquote{systemic vulnerability} consumers often face in the digital marketplace \cite{protectionepargnants}. 
By providing on-demand, actionable guidance, the system serves as a digital substitute for the advisory role traditionally held by human agents. 
One user summarized this effect: \enquote{This is more than just a Q\&A tool. It is a copilot for managing my insurance. I feel like I'm in the driver's seat for the first time.}

\subsubsection{Preference for Humans}
The qualitative data also clearly articulates the boundaries of the AI's role. A recurring theme in the comments was a strong, consistent preference for human interaction in emotionally charged situations. 
One user commented that for a \enquote{complex or litigious case} they would still prefer a human agent. 
It aligns with broader research on chatbot adoption \cite{brandtzaeg2017people,law2022effects}, which finds that users consistently prefer a \enquote{human fallback} for complex problem-solving.
This finding underscores the need to design systems with adjustable autonomy, where the agent can recognize its capabilities and seamlessly escalate the interaction to a human agent.

\subsubsection{The Critical Risk of Errors}
While most users trusted the system, the most critical theme to emerge from the feedback is the identification of significant errors by more knowledgeable participants. 
It provides concrete, real-world evidence of the system's potential to generate misinformation, a known challenge in RAG architectures \cite{gao2023,beauchemin2024quebec}. The errors fall into two categories:

\begin{enumerate}[leftmargin=*, noitemsep, topsep=0ex]
    \item \textbf{Hallucinations}: One user reported that the agent made \enquote{technical errors (like inventing an exclusion in the standard automotive insurance form in Quebec)}. 
    Another user noted a similar issue when \enquote{the agent referred to a non-standard insurance form, a completely different and very, very rare contract. It could have been misleading}. 
    These are not minor inaccuracies but fabrications of legal details that could have serious consequences.
    \item \textbf{Logical Contradictions}: The system sometimes lacked logical consistency. 
    One user experienced this directly: \enquote{The agent gave me two contradictory answers to one question. If I had relied on the first answer, I would have made a wrong decision}.
\end{enumerate}

This qualitative evidence moves the concept of \enquote{risk} from a perception measure on a Likert scale to a documented reality. 
It presents a classic challenge in real-world systems: ensuring appropriate trust calibration. 
If users over-trust the system because it is generally helpful, they may be vulnerable to its rare but critical failures. 
Therefore, designing the interaction to manage user expectations and maintain a healthy level of skepticism is as important as improving the AI's accuracy \cite{lee2004trust}.

\section{Generalization and Broader Implications}
\label{sec:generatlization}
While this study focuses on evaluating a RAG system within the specific context of Quebec insurance, its core findings offer generalizable insights for the design and deployment of conversational AI in other complex, high-stakes domains. 
We identify three key principles that transcend our use case.

First, the agent's role as a \enquote{cognitive equalizer} is a fundamental benefit of this type of system. Our results show that the system's primary value for low-literacy users was its ability to translate complex concepts into understandable language, thereby reducing cognitive load. 
More generally, this suggests that the most significant societal benefit of such AI agents may lie in their capacity to empower vulnerable populations by democratizing access to critical information, regardless of the domain in which they operate.
It is a benefit that can be replicated in any field characterized by information asymmetry and specialized jargon, such as law or healthcare. 

Second, the discovery that users value the feeling of autonomy more than knowledge acquisition itself is a crucial insight.
Users in our study were motivated not just by understanding their insurance policies, but by the tangible sense of autonomy the interaction provided. 
This shift from passive to active consumer can generalize to any domain where users face complex information asymmetry.
Moreover, an empowered decision-maker can be a universal driver of user satisfaction and adoption for AI-powered decision-support tools \cite{mcauley1989,cocca2022,yin2024}. 
Consequently, the design of future systems should prioritize features that enhance user control over purely informational content.

Third, our work highlights a universal challenge in deploying a RAG system: the need for appropriate trust calibration and human oversight. It stems from two distinct elements. First, there exists a tension between the high overall trust users have in the system and the occurrence of critical errors. 
Second, users have consistently preferred a human agent in emotionally charged scenarios. 
These issues are not unique to the insurance domain \cite{lee2004trust,brandtzaeg2017people,law2022effects}. 
These findings strongly suggest that, at least for the foreseeable future, the responsible deployment model for RAG agents in any critical domain is not full autonomy but a human-in-the-loop framework.

\section{Conclusion and Future Work}
The human-centric evaluation performed in this paper demonstrates that a sophisticated RAG architecture can successfully make complex insurance contracts understandable and empower users. 
The prototype was technically feasible and highly valued by users. 
It proved to be a powerful \enquote{cognitive equalizer}, delivering the most significant satisfaction and educational benefits to consumers with lower financial literacy. 
The system's most profound success, as perceived by users, was its ability to enhance their autonomy, shifting them from passive policyholders to active, informed participants in their insurance decisions. Beyond insurance, these findings highlight a set of broadly applicable principles for designing AI agents that can effectively and responsibly empower users in any complex information domain.
However, this study also serves as a crucial cautionary tale. 
The user-documented factual errors and logical contradictions, combined with a clear preference by users for human agents in high-stakes scenarios, confirm that the system is not yet ready for fully autonomous, unsupervised public deployment. 
In its current form, it is a powerful decision-support tool, not a replacement for human expertise.
Furthermore, while the study's conclusions are robust for a more educated, high-literacy user group, caution should be exercised when generalizing these findings to the broader population of insurance consumers.

The path forward must be guided by the limitations and preferences identified by users. 
Future work should prioritize two main avenues. 
First, enhancing trust and reliability is essential. 
It involves not only exploring advanced RAG strategies, such as self-correction to reduce factual errors, but also integrating explainability features that cite the specific, relevant sources used to generate an answer. 
Doing so requires a user-centered approach to XAI, where explanations are treated as a social and interactive process rather than a simple data dump \cite{matarese2021user}.
Second, and most critically, the system should be implemented within a robust \enquote{human-in-the-loop} framework. 
It entails developing a hybrid model with a seamless escalation path to a human expert, particularly for the complex or emotionally charged scenarios that users identified as beyond the scope of a purely automated agent. 
This approach, which leverages the strengths of both human and agentic systems through collaborative optimization, is a promising frontier for designing effective real-world applications \cite{slade2024human}.
By embracing an approach that augments rather than replaces human oversight, we can responsibly harness the power of this technology to create a more transparent and equitable insurance market for all consumers.

\begin{credits}
\subsubsection{\ackname}
This research was made possible thanks to the support of a Canadian insurance company, NSERC research grant RDCPJ 537198-18 and FRQNT doctoral research grant. We thank the reviewers for their comments regarding our work.

\subsubsection{\discintname}
Both authors have received research grants from a Canadian insurance company.

\subsubsection{Limitations}

While this study provides valuable insights into the human-centric evaluation of RAG agents, its findings should be interpreted in light of three limitations.

First, and most significantly, our participant sample was recruited from a university population and demonstrated significantly higher insurance literacy than the general Quebec population.
Consequently, the overwhelmingly positive user experience, particularly regarding ease of use and satisfaction, may not be fully generalizable to a broader, more diverse demographic, including individuals with lower educational levels or functional literacy skills who represent a key target group for such an empowering tool.

Second, the study was conducted in a controlled experimental setting with a limited 30-minute time frame. 
This environment, while necessary for data collection, does not fully replicate the real-world context of purchasing insurance or making a claim, where emotional and financial stakes are higher. 
User behavior, trust calibration, and interaction patterns may differ significantly in longitudinal, real-world deployments.

Finally, our findings are specific to the evaluated RAG architecture and the distinct legal and linguistic context of Quebec's automobile insurance system. 
The nature and frequency of system errors, as well as user preferences for human agents, could vary across different LLMs, retrieval strategies, and other complex consumer domains with distinct regulatory frameworks. 
These limitations highlight important avenues for future research, including studies with more representative populations and real deployments to assess long-term user interaction.
\end{credits}

\clearpage

\bibliographystyle{splncs04}
\bibliography{article}

\clearpage
\appendix

\section{Participants' Literacy Level Benchmark With the AMF Indice}
\label{an:amf}
To contextualize the study's user sample, we compare the participants' insurance literacy scores against the official \enquote{Indice AMF} benchmark of the Quebec population. 
\autoref{tab:sample_pop} presents a comparison of the descriptive statistics and distribution of literacy scores in our participant sample against the benchmark established by \cite{cachecho2022}. 
The analysis revealed that the participant sample demonstrated a significantly higher mean literacy score (73.29) compared to the population benchmark (54.10). 
Moreover, the \textit{t}-test results, which confirm that this difference is statistically significant (15.83) with a $p < .001$, indicate that the study's sample is not a representative cross-section of the general population but rather a more knowledgeable subgroup.
This discrepancy is likely a consequence of the convenience sampling strategy we used, which resulted in recruiting participants from a university population. 
Therefore, the subsequent evaluations of the RAG system's usability and effectiveness must be interpreted within this context. 
While the study's conclusions are robust for a more educated, high-literacy user group, caution should be exercised when generalizing these findings to the broader population of insurance consumers.

\begin{table*}
    \centering
    \begin{tabular}{@{}lccccccccc@{}}
    \toprule
               & N & Min & Max & $\mu$ & $\sigma$ & $S_k$ (Skewness)  & Kurtosis & \textit{t}-test & $p$-value \\ \midrule
    Indice AMF & 873                   & 27.78                       & 72.22                       & 54.10                    & 7.05                                   & 0.08                                         & 0.17 & \multirow{2}{*}{15.83} & \multirow{2}{*}{$< .001$}    \\
    Ours &     \participants{}                  & 33.33                       & 100.00                      & 73.29                    & 15.04                                & -0.15 & -0.56                                            \\ \bottomrule
    \end{tabular}
    \caption{Descriptive statistics of the \enquote{Indice AMF} \cite{cachecho2022} and our participant sample's literacy scores.}
    \label{tab:sample_pop}
\end{table*}

\section{UI Panel Screenshots}
\label{an:ui}
In this section, we present our UI panel screenshots.
The user's journey follows a structured flow, beginning with an authentication panel and then proceeding to a project information panel that outlines the research goals and instructions. 
The main interaction occurs within a chatbot panel, which features an interactive chat window, a conversation history, and a series of buttons for submitting conversations or requesting help. 
A key feature of the interface is the ability for users to provide granular feedback on any chatbot response by using \enquote{like/dislike} buttons and attaching specific written comments, allowing us to capture targeted qualitative insights. 

\begin{figure}[ht!]
    \centering
    \caption{Print screen of our authentication panel (in French).}
    \includegraphics[width=\linewidth]{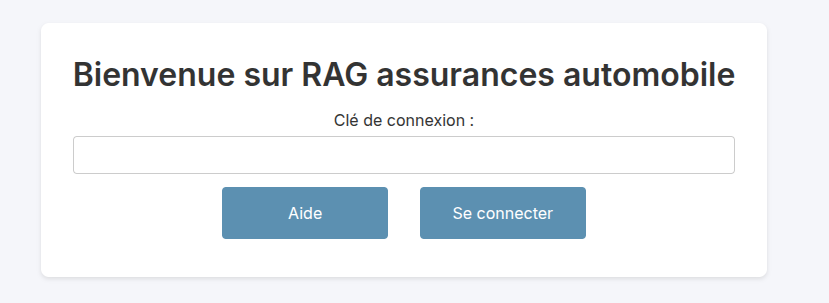}
    \label{fig:connexion}
\end{figure}

\begin{figure*}[ht!]
    \centering
    \caption{Print screen of our project details panel (in French).}
    \includegraphics[width=0.75\linewidth]{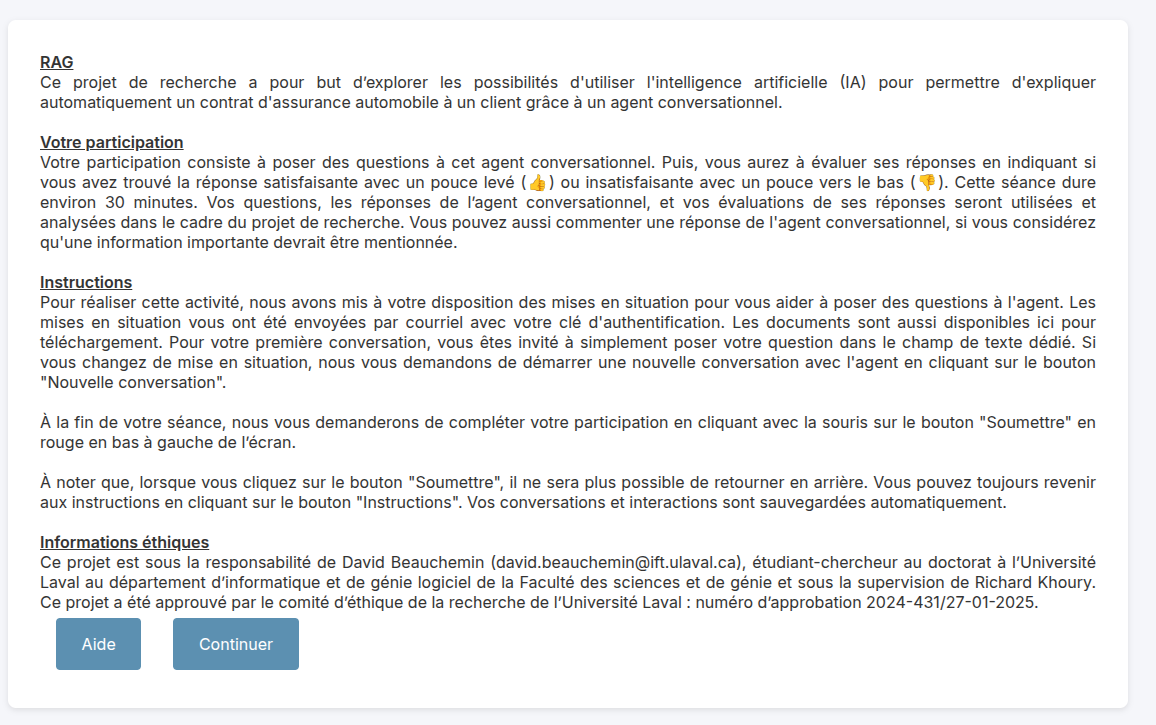}
    \label{fig:projectdetails}
\end{figure*}

\begin{figure*}[ht!]
    \centering
    \caption{Print screen of our chatbot interaction panel (in French).}
    \includegraphics[width=1\linewidth]{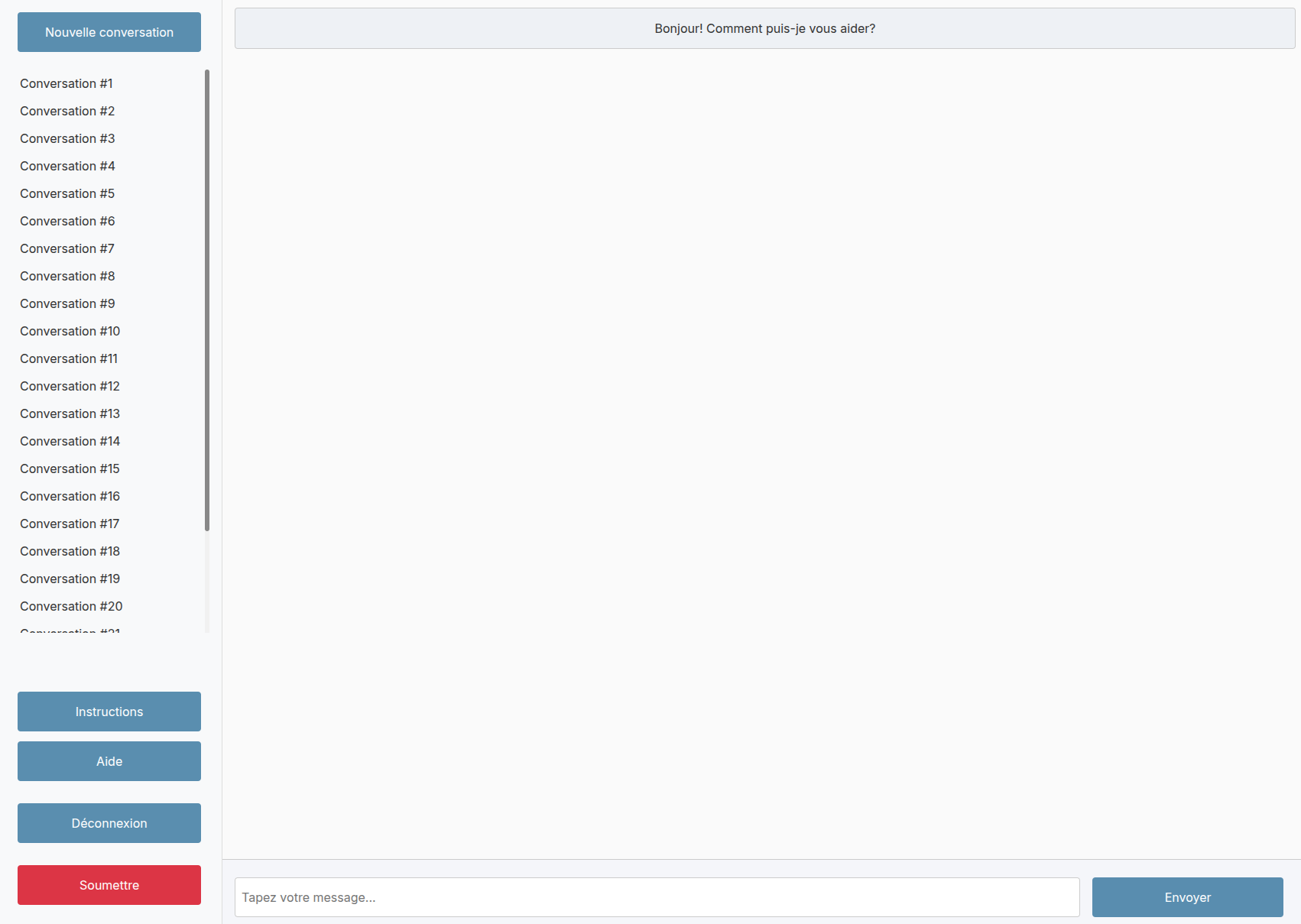}
    \label{fig:chatbot}
\end{figure*}

\begin{figure*}[ht!]
    \centering
    \caption{Print screen of an example of a conversation between a user and the chatbot.}
    \includegraphics[width=0.95\linewidth]{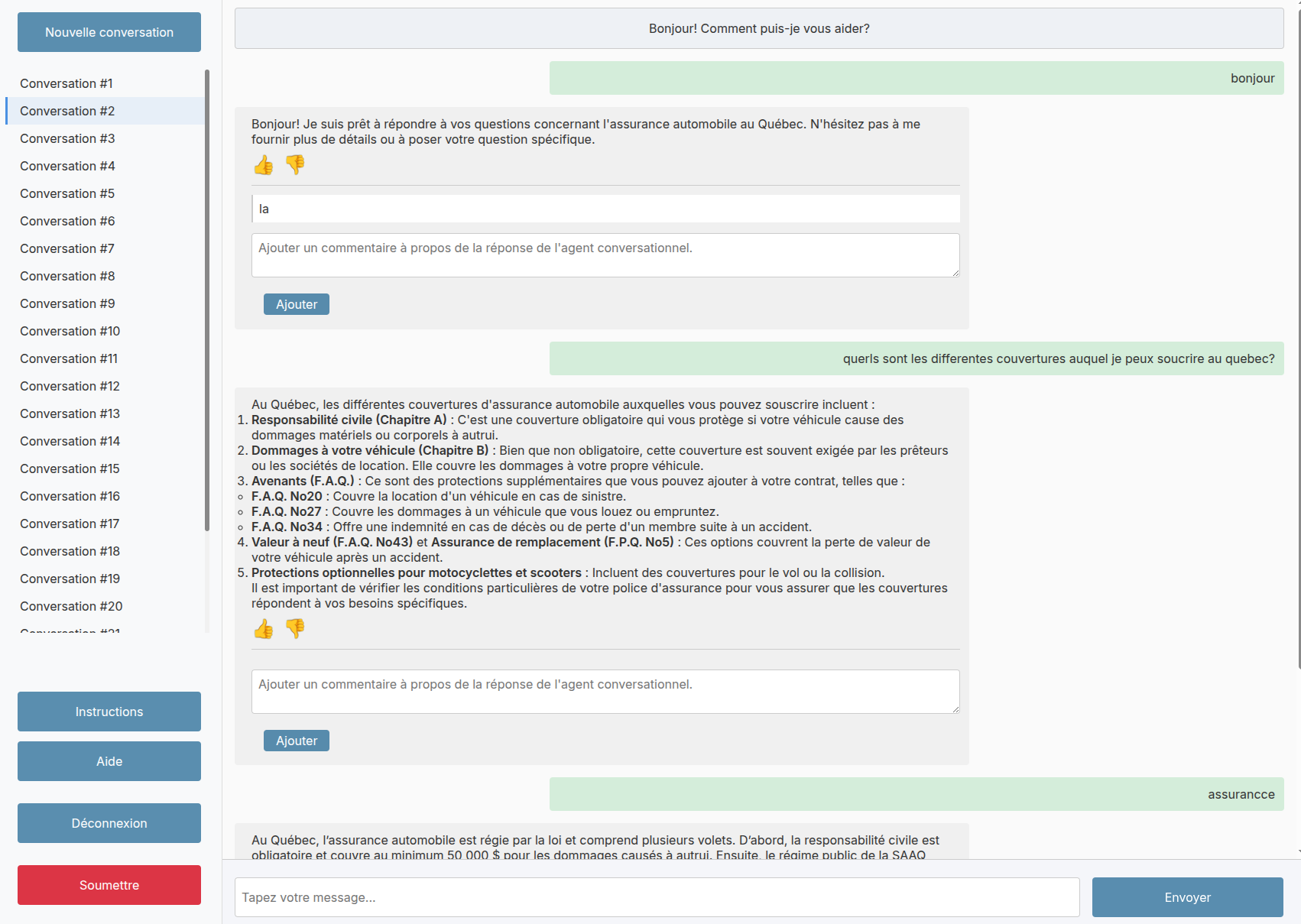}
    \label{fig:chatbotexample}
\end{figure*}

\begin{figure*}[ht!]
    \centering
    \caption{Print screen of an example of a comment under a chatbot response.}
    \includegraphics[width=0.95\linewidth]{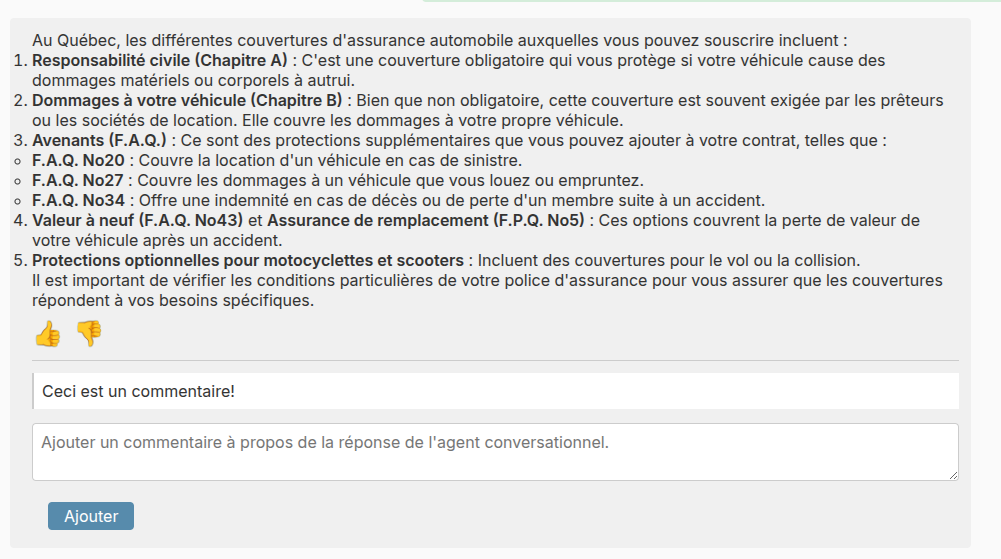}
    \label{fig:comments}
\end{figure*}

\begin{figure*}[ht!]
    \centering
    \caption{Print screen of the redirection message after a user submits their conversations.}
    \includegraphics[width=0.95\linewidth]{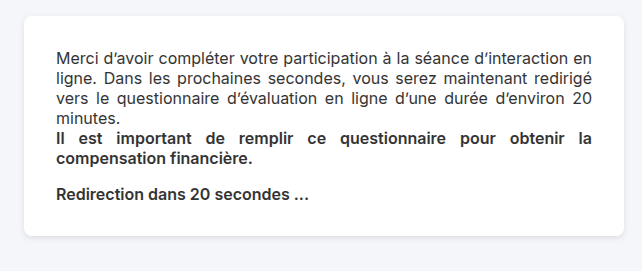}
    \label{fig:submit}
\end{figure*}

\clearpage
\section{Post-use Evaluation Survey}
\label{an:questionnaireevalen}
\begin{table*}
    \centering
    \scriptsize
    \begin{tabular}{@{}p{0.03\textwidth}p{0.06\textwidth}p{0.85\textwidth}p{0.06\textwidth}@{}}
    \toprule
        ID & Topic & Question & Scale \\\midrule
    1                           & SS   & Overall, I'm very pleased with the conversational agent.                                     & LS                          \\
    2                           & SS   & Overall, I found the conversational agent's responses accessible to read.                & LS                          \\
    3                           & SS   & Overall, I found the conversational agent's answers easy to understand.          & LS                          \\
    4                           & SS   & Overall, I found that the conversational agent's responses were well-explained.         & LS                          \\
    5                           & SS   & Overall, my interaction with the conversational agent was very satisfactory.                 & LS                          \\
    6                           & SS   & The conversational agent performed satisfactorily: it did not generate any nonsensical responses, nor did it exhibit any failures or bugs.                                                                                                                                                                                                                                                                                                                     & LS                          \\
    7                           & SS   & I trust the conversational agent to give me reliable answers.                        & LS                          \\
    8                           & SS   & I'd like to continue using this type of solution (i.e. a conversational agent) in the future.   & LS                          \\
    9                           & SS   & Overall, I enjoyed interacting with the conversational agent.                                  & LS                          \\
    10                          & SS   & Overall, I was able to improve my understanding of car insurance products thanks to the conversational agent.                                                                                                                                                                                                                                                                                                                                              & LS                          \\
    11                          & SS   & Explain how the conversational agent helped or hindered your understanding of car insurance products.                                                                                                                                                                                                                                                                                                                                         & Text              \\
    12                          & CE             & Overall, I needed much concentration to use the conversational agent.     & LS                          \\
    13                          & CE            & Overall, I needed much concentration to understand the conversational agent's answers.                                                                                                                                                                                                                                                                                                                                                   & LS                          \\
    14                          & CE            & Overall, I needed to make a sustained mental effort to use the conversational agent.                                                                                                                                                                                                                                           & LS                          \\
    15                          & CE            & Overall, I needed to make a sustained mental effort to understand the conversational agent's responses.                                                                                                                                                                                                                                                                                                                                              & LS                          \\
    16                          & PA                  & I feel that this type of solution (i.e. a conversational agent) has the potential to increase my autonomy when purchasing, renewing or making a claim on my insurance policy.                                                                                                                                                                                                                                           & LS                          \\
    17                          & PA                  & I have the impression that this type of solution (i.e. a conversational agent) has the potential to help other people become self-sufficient when buying, renewing their insurance policy or making a claim.                                                                                                                                                                                                                                             & LS                          \\
    18                          & PR                    & I see a risk in using this type of system (i.e. a conversational agent). & LS                          \\
    19                          & PR                    & I see a risk for society in using this type of system (i.e. a conversational agent).  & LS                          \\
    20                          &       PR       & I'd like to use this type of digital conversational agent specialized in property and casualty insurance to get reliable answers to my questions, especially when: purchasing an auto and/or home insurance policy; renewing insurance; cancelling insurance; making changes to my insurance file or policy; making a claim following a loss; none of the above. & MP \\
    21                          &                & Do you have any comments?                                                                             & Text             \\ \bottomrule
    \end{tabular}
        \caption{Our translated post-use evaluation survey questions designed to evaluate our agent for our insurance use-case. \enquote{LS} refers to the Likert scale presented in \autoref{fig:legend}, while \enquote{MC} refers to \enquote{Multiple-choices}.}
    \label{tab:questionnaireevalen}
\end{table*}

\end{document}